\documentclass[aps,preprint,nofootinbib]{revtex4-1}
\usepackage{graphicx}
\begin{document}
\title{\Large \bf Type D Solutions of 3D New Massive Gravity }
\author{\large Haji Ahmedov and Alikram N. Aliev}
\address{Feza G\"ursey Institute, \c Cengelk\" oy, 34684   Istanbul, Turkey}
\date{\today}
\begin{abstract}

In a recent reformulation of three-dimensional new massive gravity (NMG), the field equations of the theory consist of a massive (tensorial) Klein-Gordon type equation with a curvature-squared source term and a constraint equation. Using this framework, we present {\it all algebraic type D solutions} of NMG with constant and nonconstant scalar curvatures. For constant scalar curvature, they include homogeneous anisotropic solutions which encompass  both solutions originating from  topologically massive gravity (TMG), Bianchi types $ II $\,,  $ VIII $\,, $ IX $, and those of  non-TMG origin,  Bianchi types $ VI_{0} $ and $ VII_{0} $. For  a special relation between the cosmological and mass parameters, $\lambda=m^2 $, they also include conformally flat solutions, and in particular those being locally isometric to the previously-known  Kaluza-Klein  type $ AdS_2\times S^1 $ or $ dS_2\times S^1 $ solutions. For nonconstant scalar curvature, all the solutions are conformally flat and exist only  for $\lambda=m^2 $. We find two general metrics which possess at least one Killing  vector and comprise all such solutions. We also discuss some properties of these solutions, delineating among them black hole type solutions.

\vspace{8mm}

PACS numbers: 04.60.Kz, 11.15.Wx

\vspace{8mm}

\begin{center}
{\it Dedicated to the  memory of Yavuz Nutku}
\end{center}

\end{abstract}

\maketitle

\section{Introduction}

General relativity in  three dimensions  suffers from  a number of barren properties: The usual Newtonian limit does not exist, there are no local dynamical degrees of freedom and no  black hole solutions without a negative cosmological constant \cite{star, djh} (see  also \cite{carlip} and references therein). With a negative cosmological constant, the theory admits black holes with an asymptotically anti-de Sitter (AdS$_3$) behavior, which are known as Banados-Teitelboim-Zanelli (BTZ)  black holes \cite{btz}. The BTZ black holes  have played a profound  role in understanding  the classical and quantum structures  of the ``simple"  general relativity, albeit its final quantum status continues to remain unclear \cite{witt}. On par with this, intriguing developments have also been towards giving the theory dynamical degrees of freedom by adding higher-derivative terms to the Einstein-Hilbert (EH) action. Such a  procedure generically introduces a topological mass parameter in the theory and  entails  the appearance of propagating modes with negative energy, known as {\it ghosts}. However, there are some fortunate exceptions in which cases the ghosts can be avoided. Among such exceptions, topologically massive gravity (TMG) occupies a central place \cite{djt, deser}. This theory is obtained by adding to the EH action a gravitational  Chern-Simons term which is third-order in a derivative expansion.  Nevertheless, the particle content of the theory is revealed in a usual way, by linearization of the field equations about a flat Minkowski spacetime. The resulting wave equation represents  a single massive spin 2 mode of helicity either $+$2 or $-$2, depending on the sign of the topological mass parameter. This mode has positive energy only for the opposite sign (with respect to the conventional one  in four dimensions) of the EH term in the total action. Thus, TMG  is a  unitary theory only for the ``wrong" sign of the EH term and it is  a parity-violating theory by its very nature \cite{djt}.

Recently, it was shown that there is a special case of TMG, determined by a critical value of the topological mass parameter, for which one can still allow the usual ``right"  sign of the EH term \cite{strom1}. In this case, the massive gravitons disappear and the theory admits BTZ black holes with positive mass. This in turn resolves the unitarity conflict between the bulk and  boundary theories in the context of AdS$_3$/CFT$_2$ correspondence.
That is, the theory of TMG in its special case becomes dual to a conformal field theory (CFT$_2 $) on the boundary (see Refs.\cite{strom1, strom2} for details). This result has also renewed the interest in massive gravity theories in three dimensions on their own right and, as another fortunate exception, a new ghost-free theory of massive gravity appeared \cite{bht}. This theory is known as {\it new massive gravity }(NMG). It is a  parity-preserving theory and is obtained by adding to the EH action a particular quadratic-curvature term. Though such an action gives rise to the field equations of fourth-order in derivatives, its canonical structure  involves two propagating  massive modes of helicities $ \pm 2 $. Moreover, the equivalence of this action, when expanding about the Minkowski vacuum, to the Fierz-Pauli action confirms unitary property of the theory. Subsequently, this property  has also been confirmed in  \cite{naka, deser1, tekin}.

Further developments have been towards searching for exact solutions to the theory of NMG and studying its holographic properties. Shortly after the advent of NMG, it was found that the theory admits regular warped AdS$_3$ black holes which are counterparts of those appearing in TMG \cite{clement}. Continuing the line of analogy with TMG solutions \cite{nb, nutku, gm}, the other constant scalar curvature solutions of NMG, such as AdS$_3$-waves \cite{giri1} and Bianchi type IX homogeneous space solution \cite{baka} were also found. A new class of solutions with nonconstant scalar curvature, which includes the black hole, gravitational soliton and kink type solutions, was discussed in \cite{berg, troncoso}. Other studies of the issue of exact solutions to NMG can be found in  \cite{giri2, mg, gho, nam, sot1}. The holographic properties  of NMG exhibit some similarities with  those of TMG: The bulk/boundary unitarity conflict still exists and its sensible resolution at some  ``chiral" points  turns out to depend crucially on the nature of asymptotic boundary conditions \cite{berg, liu1,liu2, sinha}. Therefore, finding pertinent stable vacua of the theory is of great importance and this  motivates us to search for further exact solutions.

In striving to undertake an exhaustive investigation of the exact solutions, the theory of NMG was recently reformulated in terms of a first-order differential operator (appearing in TMG and resembling a Dirac type operator) and the traceless Ricci tensor \cite{ah1}. With this novel proposal, the field equations of NMG consist of a massive (tensorial) Klein-Gordon type equation with an ``effective"  curvature-squared  source term and a constraint equation. This approach  has a striking consequence for finding   algebraic types D and N (in the Petrov-Segre classification) exact solutions to NMG. Remarkably, for these types of spacetimes, possessing constant scalar curvature, one can interpret  TMG as the  ``square root"  of NMG. This allows us to establish  a simple framework for mapping of all types D and N solutions of TMG into NMG \cite{ah1}. In other words, there is no need to solve  in each case the field equations of NMG for a given metric ansatz as it becomes possible to generate solutions to NMG  from the known solutions of TMG. Moreover, the novel proposal  turns out to be a powerful tool for finding new  types D and N solutions to NMG which do not have their counterparts in TMG. Some intriguing and  simple examples  of  these ``non-TMG origin"  solutions were given \cite{ah1}. In the meantime, a universal metric for all type N solutions of NMG was found in \cite{ah2}.

The purpose of this paper is to present {\it all algebraic type D solutions}  of NMG with both constant and nonconstant curvatures. In Sec.II we begin by describing a theoretical framework which includes reformulation of NMG in terms of a first-order differential operator, appearing TMG  and  acting on the traceless Ricci tensor. In Sec.III  we introduce an orthonormal  basis  of three real vectors (one timelike and  two spacelike vectors) and  describe the properties of type D  spacetimes in  NMG, in terms of these vectors and their covariant derivatives. Here we also prove two defining mathematical  statements for these spacetimes. In Sec.IV we present the exact solutions with constant scalar curvature, which include homogeneous  anisotropic space solutions of both TMG  and  non-TMG origins. Next, we  present conformally  flat solutions  with constant scalar curvature, for which the cosmological and mass parameters are related by $ \lambda = m^2 $.  In Sec.V  we discuss the solutions with nonconstant scalar curvature and possessing at least one Killing  vector. They all are conformally flat and require  $ \lambda = m^2 $ by their very existence. We find two general metrics which comprise  all such solutions. We also discuss some properties of these solutions, among them are black hole type solutions.

\section{Theoretical Framework}

For our purposes in the following it is convenient to begin with TMG. The theory is described by the action \cite{djt,deser}
\begin{eqnarray}
S &=& \frac{1}{16\pi G} \int d^3x \sqrt{-g}\left(R - 2 \Lambda +\frac{1}{\mu}\, {\cal{L}}_{CS}\right)\,,
\label{tmgaction}
\end{eqnarray}
where $ R $ is the three-dimensional Ricci scalar, $ \Lambda $ is the cosmological constant, $ \mu $ is the  mass parameter and the Chern-Simons term is given by
\begin{eqnarray}
{\cal{L}}_{CS} &=& \frac{1}{2}\, \epsilon^{\,\mu\nu\rho} \Gamma^{\alpha}_{\mu \beta}\left(\partial_{\nu}\Gamma^{\beta}_{\rho \alpha}+ \frac{2}{3}\,\Gamma^{\beta}_{\nu \gamma}\Gamma^{\gamma}_{\rho \alpha}\right)
\,.
\label{nmgaction}
\end{eqnarray}
Here $ \epsilon^{\,\mu\nu\rho} $ is the  Levi-Civita tensor and the quantities $ \Gamma^{\mu}_{\alpha \beta} $ are the usual Christoffel symbols.

The field equations of TMG, obtained from this action, are given by
 \begin{eqnarray}
G_{\mu\nu}  + \Lambda g_{\mu\nu} + \frac{1}{\mu}\, C_{\mu\nu}=0\,,
\label{tmgfieldeqs1}
\end{eqnarray}
where $\, G_{\mu\nu}= R_{\mu\nu} - \frac{1}{2}\, g_{\mu\nu} R \,$ is the Einstein tensor and  $ C_{\mu\nu} $ is the Cotton tensor given by
\begin{eqnarray}
C_{\mu\nu} &= & {\epsilon_{\mu}}^{\alpha\beta}\left(R_{\nu\beta} - \frac{1}{4}\, g_{\nu\beta} R\right)_{;\alpha}
\label{cotton}
\end{eqnarray}
and the semicolon stands for covariant differentiation. We note that this tensor is a symmetric, traceless and covariantly conserved quantity.

It is not difficult to see that with the traceless Ricci tensor
\begin{eqnarray}
S_{\mu\nu}= R_{\mu\nu}-\frac{1}{3}\,g_{\mu\nu} R\,,
\label{trlessricci}
\end{eqnarray}
 equation (\ref{tmgfieldeqs1}) can be put in the form
\begin{eqnarray}
S_{\mu\nu} + \frac{1}{\mu}\, C_{\mu\nu}=0\,.
\label{tmgfieldeqs2}
\end{eqnarray}

Next, we introduce a first-order differential operator   $ \,{D\hskip -.25truecm \slash}\, $ which is  defined as
\begin{eqnarray}
{D\hskip -.25truecm \slash}\,\Phi_{\mu\nu} &= & \frac{1}{2}\left( {\epsilon_{\mu}}^{\alpha\beta} \Phi_{\nu\alpha ;\beta} + {\epsilon_{\nu}}^{\alpha\beta} \Phi_{\mu\alpha ;\beta} \right),
\label{doper1}
\end{eqnarray}
where  $ \Phi_{\mu \nu} $  is a symmetric tensor. It is straightforward to show that  with the subsidiary relation
\begin{eqnarray}
{\Phi_{\mu\nu}}^{;\nu} &= & (\Phi_{\nu}^{\,\,\nu})_{;\mu} \,,
\label{cond1}
\end{eqnarray}
equation (\ref{doper1}) takes the most simple form
\begin{eqnarray}
{D\hskip -.25truecm \slash\,}\Phi_{\mu\nu} &= & {\epsilon_{\mu}}^{\alpha\beta} \Phi_{\nu\alpha ;\beta}\,.
\label{doper2}
\end{eqnarray}
Since  condition (\ref{cond1}) is fulfilled for the Schouten tensor
\begin{eqnarray}
\Phi_{\mu\nu}= R_{\mu\nu}-\frac{1}{4}\,g_{\mu\nu} R\,,
\label{phitensor}
\end{eqnarray}
it is easy to verify that  \begin{eqnarray}
C_{\mu\nu} &= & -{D\hskip -.25truecm \slash\, }  \Phi_{\mu\nu}= - {D\hskip -.25truecm \slash\, } S_{\mu\nu}\,.
\label{cot1}
\end{eqnarray}
With these relations the field equations of TMG, as seen from equation (\ref{tmgfieldeqs2}), acquire the form
\begin{eqnarray}
{D\hskip -.25truecm \slash\,} S_{\mu\nu} & = &  \mu S_{\mu\nu}\,.
\label{tmgdirac}
\end{eqnarray}
We note that this equation resembles the Dirac equation $ {D\hskip -.25truecm \slash\, }\Psi_{A}=\gamma^{\,\mu B}_{A} \nabla_{\mu} \Psi_B = \mu \Psi_{A} $ for a massive spinor field. In the case under consideration, the operator $ {D\hskip -.25truecm \slash\,} $ acts on the  traceless Ricci tensor, $ {D\hskip -.25truecm \slash\,} S _{\mu\nu} =D_{(\mu\nu)}^{\,\,\,(\alpha\beta) \rho} \nabla_{\rho} S_{\alpha\beta }$,  and therefore  it looks like the ``cousin" of the Dirac operator (see also Refs.\cite{ah1, ah2}).

Using equation  (\ref{tmgdirac}), it easy to show that the secondary action of  the operator  $ {D\hskip -.25truecm \slash} $  yields the  Klein-Gordon type equation
\begin{eqnarray}
\left({D\hskip -.25truecm \slash\,}^2  - \mu^2 \right)S_{\mu\nu} & = &  0\,.
\label{tmgkg}
\end{eqnarray}
It is  also important to note that with the differential operator $ {D\hskip -.25truecm \slash} \,$ we  have the following identity
\begin{eqnarray}
({D\hskip -.25truecm \slash\,}^2  S_{\mu\nu})^{;\nu}& = & \epsilon_{\mu}^{ \ \  \rho\sigma}
 S_{\rho\nu} \ C^{\nu}_{\  \sigma}\,.
 \label{id1}
\end{eqnarray}
This identity can be verified  by using the relation
\begin{eqnarray}
{D\hskip -.25truecm \slash\,}^2  S_{\mu\nu} &=& {\Phi_{\mu\nu ; \sigma}}^{;\sigma} - {\Phi_{\mu\rho ;\nu}}^{;\rho}
\label{nmgdirac2}
\end{eqnarray}
along with the fact that in three dimensions the Riemann tensor is given as
\begin{eqnarray}
R_{\mu \nu \alpha \beta} & = & 2\left(R_{\mu [\alpha}g_{\beta]\nu} - R_{\nu [\alpha} g_{\beta]\mu}
-\frac{R}{2}\, g_{\mu[\alpha}g_{\beta]\nu}\right),
\label{rimtoric}
\end{eqnarray}
where the square brackets stand for antisymmetrization over the indices enclosed.

There also exists another important identity
\begin{eqnarray}
({D\hskip -.25truecm \slash\,}^2  S_{\mu\nu})^{;\mu ;\nu}& = & C_{\mu\nu}
C^{\mu\nu} - S^{\mu\nu} {D\hskip -.25truecm \slash\,}^2
S_{\mu\nu}\,,
\label{id2}
\end{eqnarray}
that is obtained  by  a straightforward calculation  of the quantity  $ S^{\mu\nu} {D\hskip -.25truecm \slash\,}^2 S_{\mu\nu} $, using  the definition in  (\ref{doper1}), with equations (\ref{doper2})  and (\ref{cot1}) in mind, as well as  the identity in (\ref{id1}).

We now proceed to  NMG whose action has the form \cite{bht}
\begin{eqnarray}
S &=& \frac{1}{16\pi G} \int d^3x \sqrt{-g}\left(R - 2 \lambda -\frac{1}{m^2}\, K \right),
\label{nmgaction}
\end{eqnarray}
where $\lambda $ is the cosmological parameter, $ m $ is the mass parameter and the scalar  $ K $ is given by
\begin{eqnarray}
K &=& R_{\mu\nu}R^{\mu\nu}- \frac{3}{8} \,R^2\,.
\label{k}
\end{eqnarray}
In a recent work \cite{ah1}, it was shown that the  field equations, which follow from this action, can be written in terms of the square of the operator  $ {D\hskip -.25truecm \slash} \,$,  resulting in  the Klein-Gordon type equation
\begin{eqnarray}
\left({D\hskip -.25truecm \slash\,}^2  - m^2 \right) S_{\mu\nu} & = & T_{\mu\nu}\,,
\label{nmgfieldeqs2}
\end{eqnarray}
where $ T_{\mu\nu} $  can be thought of as the  effective source tensor. We have  
\begin{eqnarray}
T_{\mu\nu}& = &  S_{\mu\rho}S^{\rho}_{\,\nu}- \frac{R}{12}\,S_{\mu\nu}- \frac{1}{3}\,g_{\mu\nu} S_{\alpha\beta}S^{\alpha\beta}\,.
\label{source}
\end{eqnarray}
Equation (\ref{nmgfieldeqs2}) is also accompanied by  the subsidiary (constraint) equation
\begin{eqnarray}
S_{\mu\nu}S^{\mu\nu} + m^2 R - \frac{R^2}{24} & = &  6 m^2 \lambda\,.
\label{newtrfeqs}
\end{eqnarray}
It is remarkable that such a description of NMG  greatly simplifies the search for exact solutions to the theory. It is instructive to demonstrate this using the case of maximally symmetric  solutions to TMG, for which the Cotton tensor in (\ref{tmgfieldeqs2}) vanishes  identically and we have $ S_{\mu\nu}=0 $. Then equation (\ref{nmgfieldeqs2}) is trivially satisfied, whereas equation (\ref{newtrfeqs}) yields
\begin{eqnarray}
\Lambda= 2 m^2\left(1\pm \sqrt{1-\lambda/m^2}\right),
\label{maxsym}
\end{eqnarray}
where we have used the fact that  $ R = 6 \Lambda $. Thus, with the cosmological constant adjusted according to this relation, every maximally symmetric solution of TMG (including its identification with a BTZ  black hole \cite{btz}) can be  mapped into two inequivalent solutions of NMG \cite{bht,clement}.

The most striking feature of this description emerges for algebraic types D and N spacetimes. It is straightforward to verify that for these spacetimes the following relation
\begin{eqnarray}
T_{\mu\nu} & = & \kappa  S_{\mu\nu}\,
\label{geocond}
\end{eqnarray}
holds, where  $ \kappa  $ is  a function of the scalar curvature. With this relation, equation (\ref{nmgfieldeqs2}) reduces to the form
\begin{eqnarray}
\left({D\hskip -.25truecm \slash\,}{^2}-  \mu^2\right)S_{\mu\nu} & = &  0\,,
\label{kgfinal0}
\end{eqnarray}
where
\begin{eqnarray}
 \mu^2 & = &  m^2 + \kappa\,.
\label{kgfinal}
\end{eqnarray}
We see that for type N spacetimes and type D spacetimes with constant scalar curvature this equation is equivalent to that given (\ref{tmgkg}). Thus, comparing equations (\ref{tmgdirac}), (\ref{tmgkg}) and (\ref{kgfinal0})
we conclude that for these types of spacetimes  TMG can be interpreted as the square root of NMG.  This is a remarkable fact and it paves the way for mapping all types D and N  exact solutions of TMG  into NMG  by means of an algebraic procedure that adjusts the physical parameters of the corresponding solutions in both theories \cite{ah1}. On the other hand, it is clear that equation (\ref{kgfinal0})  admits types D and N  exact solutions  of non-TMG  origin  as well. (The square root does not always exist!) Some simple examples of such  solutions  were given in \cite{ah1}. Meanwhile, in \cite{ah2} we managed to find a universal metric which  comprises all  type N solutions of NMG.  Here we wish  to continue the line of  works \cite{ah1} and  \cite{ah2}, giving an exhaustive investigation of all type D solutions of NMG.

\section{Type D spacetimes}

Let us introduce an orthonormal  basis  of three real vectors $ \{t_\mu, \ s_\mu, \ v_\mu \}$, satisfying the relations
\begin{equation}
t_\mu t^\mu =-1\,, ~~~~s_\mu s^\mu =1\,,~~~~v_\mu v^\mu =1\,,
\label{ortbasis}
\end{equation}
with all other products vanishing, such that  the spacetime metric  can be written in the form
\begin{equation}
g_{\mu\nu}= -t_\mu t_\nu +s_\mu s_\nu+v_\mu v_\nu\,.
\label{metric}
\end{equation}
In three dimensions, depending on whether the one-dimensional eigenspace of  $ S^{\mu}_{\,\,\nu} $ is timelike or spacelike,  one can distinguish types $ D_t $ and $ D_s $ spacetimes, respectively \cite{chow1} (see also \cite{hall, garcia}). For type $ D_t $ spacetime, the canonical form of  the traceless Ricci tensor is given by
\begin{eqnarray}
S_{\mu\nu} & = & p \left(g_{\mu\nu} +3 t_{\mu} t_{\nu}\right),
\label{riccican}
\end{eqnarray}
whereas, for type $ D_s $ spacetime we have
\begin{eqnarray}
S_{\mu\nu} & = & p \left(g_{\mu\nu} - 3 s_{\mu} s_{\nu}\right),
\label{riccicans}
\end{eqnarray}
where $ p $ is a scalar function. In what follows, we shall perform  a detailed analysis for type $ D_t $  spacetimes. The same analysis can be performed for type $ D_s $  spacetimes as well, either directly or simply by an appropriate analytical continuation  of the type $ D_t $  results.

Clearly, using expression (\ref{riccican}) in equation (\ref{newtrfeqs}) one can express  the function $ p $ in terms of the  scalar curvature, the mass and cosmological parameters, $ m $ and $ \lambda $. Thus, we have
\begin{eqnarray}
6p^2 &= & 6 m^2 \lambda - m^2 R + \frac{R^2}{24}\,.
\label{al1}
\end{eqnarray}
On the other hand, taking into account  expression (\ref{riccican}) in equation (\ref{geocond}) and  using the result in equation (\ref{kgfinal}), we find that
\begin{eqnarray}
 \mu^2 & = &  m^2 - p - \frac{R}{12}\,.
\label{kgfinal1}
\end{eqnarray}
We turn now to the Cotton tensor in (\ref{cotton}). It is straightforward  to verify  that the most general form of this tensor, written in terms of the orthonormal basis vectors, must have the following representation
\begin{eqnarray}
C_{\mu\nu} &=& a S_{\mu\nu}+ b \left( s_\mu s_\nu - v_\mu v_\nu \right) +   2 c \, v_{(\mu} s_{\nu )} + 2 h \,t_{(\mu}s_{\nu)}\,,
\label{cotrepre1}
\end{eqnarray}
where the round brackets denote symmetrization of the indices enclosed. The coefficients $ a $, $ b $, $ c  $ and $ h $  are scalar  functions and  we have gauged out the term proportional to  $ t_{(\mu} v_{\nu )}$ using the invariance of equations (\ref{metric}) and (\ref{riccican}) with respect to rotations in the $( s_\mu, v_\mu )$-plane. Using this representation in  equation (\ref{id1}), we find that
\begin{eqnarray}
({D\hskip -.25truecm \slash\,}^2  S_{\mu\nu})^{;\nu}& = & 3 p h v_\mu\,.
 \label{id22}
\end{eqnarray}
Next, taking the divergence of equation (\ref{kgfinal0}) and  comparing the result with  equation (\ref{id22}), we arrive at the relation
\begin{eqnarray}
{S_{\alpha}}^{\beta} (\mu^2)_{; \beta} + \frac{1}{6}\,\mu^2 R_{; \alpha}
& = &  3 p  h  v_\alpha\,,
\label{id3}
\end{eqnarray}
where we have used the  contracted Bianchi  identity
\begin{equation}
{S_{\mu\nu}}^{ ;\nu}=\frac{1}{6}\, R_{;\mu}\,.
\label{contrbian}
\end{equation}
Further it is convenient to  consider in (\ref{kgfinal1})
the cases of $\mu^2 \neq 0 $ and $\mu^2 =0 $ separately: First, we  begin with the case $\mu^2 \neq 0 $. Contracting both sides of equation (\ref{id3}) with $ s^\mu $, we find that
\begin{eqnarray}
s^\mu R_{;\mu} &=& 0\,.
 \label{sR}
\end{eqnarray}
That is, the scalar curvature is constant along the basis vector $ s^\mu $. Meanwhile, the contraction of equation (\ref{id3}) with $ v^\mu $ yields
\begin{eqnarray}
h & = & \frac{\mu^2}{12p}\, v^\mu  R_{; \mu}\,,
\label{h}
\end{eqnarray}
where we have used equations (\ref{al1}) and (\ref{kgfinal1}). We note that the same contraction procedure with $ t^\mu $  gives a trivial result. Next, we  need to find the representation for  the covariant derivative of the vector $ t^\mu $. For this purpose, we substitute expression (\ref{riccican}) into  equation (\ref{cot1}) and  taking into account the relations
\begin{eqnarray}
t_{\mu} &=& \epsilon_{\mu\nu \rho} s^{\nu} v^{\rho},~~~~ s_{\mu} = \epsilon_{\mu\nu \rho} t^{\nu} v^{\rho},~~~~ v_{\mu} = \epsilon_{\mu\nu \rho} s^{\nu} t^{\rho}
\label{orient1}
\end{eqnarray}
together with  the Bianchi identity  in (\ref{contrbian}), we compare the result with the representation of the Cotton tensor given in (\ref{cotrepre1}). After  straightforward calculations, we find that
\begin{eqnarray}
t_{\mu;\nu} & = & \frac{a}{3} \,\epsilon_{\mu\nu}^{\ \ \ \rho}\,t_\rho - \frac{2b}{3p} \, s_{(\mu}v_{\nu)} + \frac{c}{3p}\,\left(s_\mu s_\nu-v_\mu v_\nu\right)
+ \frac{g}{3p} \,\left(s_\mu s_\nu + v_\mu v_\nu \right) +  \frac{f}{3p}\, v_\mu t_\nu\,,
\label{covart}
\end{eqnarray}
where
\begin{eqnarray}
\label{gdef}
g&=&  \frac{3 p}{2}\,{t^\mu}_{;\mu}= - \left(p{\,^\prime} + \frac{1}{12}\right) t^\mu R_{;\mu} \,,\\
f &=&  \left(p{\,^\prime} -\frac{1}{6}\right)  v^\mu R_{;\mu}\,,
\label{fdef}
\end{eqnarray}
the prime stands for the derivative with respect to the scalar curvature. In obtaining (\ref{gdef}) we have also used expression (\ref{riccican}) in equation (\ref{contrbian}) with the subsequent contraction of the result with $ t^\mu $.

We are now ready to prove that the following statement: {\it Suppose that the scalar function $ h = 0 $. Then,  for  $ \mu^2 \neq 0 $, all type D solutions of  NMG must possess constant scalar curvature}.

The proof will be given by contradiction, assuming that  $ R \neq const $. First, we note that  for $ h = 0 $  equation (\ref{h}) yields
\begin{eqnarray}
v^\mu R_{;\mu} &=& 0\,.
 \label{vR}
\end{eqnarray}
Moreover, comparing  equations  (\ref{id1}) and  (\ref{id22}) we  obtain that
\begin{eqnarray}
C_{\rho[\mu} S_{\nu]}^{~~\rho} &=& 0\,.
 \label{georel}
\end{eqnarray}
This relation is trivially satisfied in the TMG case, whereas it is a governing geometrical relation for the NMG solutions. It is important to note that with $ h = 0 $  in (\ref{cotrepre1}), one can always gauge out either $ b $  or $ c $ due to the rotational symmetry of the spacetime metric (\ref{metric}) in the $( s_\mu, v_\mu )$-plane. Therefore, in the following we can set $ c $ equal to zero, without loss of generality.

We also note that  with relations (\ref{sR})  and (\ref{vR}), one can  always write down the simple decomposition
\begin{eqnarray}
R_{; \mu} &=&  \chi \,t_\mu\,,
 \label{Rdecomp}
\end{eqnarray}
where $ \chi $ is some function and $ \chi \neq 0 $ as  $ R \neq const $ by our assumption. Using the fact that $ R_{[; \mu ; \nu]}= 0 $, we take the covariant derivative of this equation. As a consequence, we have
\begin{eqnarray}
\chi\, t_{[\mu ; \nu]} + t_{[\mu} \chi_{ ;\nu]} &=& 0 \,.
 \label{chieq}
\end{eqnarray}
Substituting now   (\ref{covart}) into this equation  and contracting the result with $ v^{\mu} s^{\nu} $,  we  find that $ a=0 $, in addition  to $ c=0 $. Thus, equation (\ref{covart}) reduces  to the form
\begin{eqnarray}
t_{\mu;\nu} & = &  - \frac{2b}{3p} \, s_{(\mu}v_{\nu)} + \frac{g}{3p} \,\left(s_\mu s_\nu + v_\mu v_\nu \right).
\label{covart1}
\end{eqnarray}
Meanwhile, the  Cotton tensor in (\ref{cotrepre1}) becomes as
\begin{eqnarray}
C_{\mu\nu} &=&  b \left( s_\mu s_\nu - v_\mu v_\nu \right)\,.
\label{cotrepre2}
\end{eqnarray}
The value of $ b $ in these equations can be fixed by using (\ref{id2}) along with (\ref{kgfinal0}) and (\ref{id22}). We find that
\begin{eqnarray}
b^2 &=& 3 p^2 \mu^2 .
 \label{bvalue}
\end{eqnarray}
Next, we calculate the divergence of expression (\ref{cotrepre2}). Taking into account the property $ {C_{\mu\nu}}^{;\nu}= 0 $,  and  contracting the  result with the vectors $ s^{\mu} $ and $ v^{\mu} $, we have
\begin{eqnarray}
{s_\mu}^{;\mu} = 0 \,,~~~~~{v_\mu}^{;\mu} = 0 \,,
\label{divvso}
\end{eqnarray}
where we have also used  relations (\ref{sR})  and (\ref{vR}).

With  equations (\ref{covart1}) and (\ref{divvso}), as well as   using the properties of the orthonormal basis vectors given in (\ref{ortbasis}), it is  straightforward to write down the explicit expressions for the covariant derivatives of the basis vectors $ s_{\mu} $ and $ v_{\mu} $. They are given by
\begin{eqnarray}
\label{scovar}
s_{\mu;\nu} & = &  - \frac{b}{3p} \, t_{\mu} v_{\nu} + \frac{g}{3p} \, t_\mu s_\nu +  \zeta v_\mu t_\nu \,,\\[2mm]
v_{\mu;\nu} & = &  - \frac{b}{3p} \, t_{\mu} s_{\nu} + \frac{g}{3p} \, t_\mu v_\nu - \zeta s_\mu t_\nu \,,
\label{vcovar}
\end{eqnarray}
where $ \zeta $ is a scalar function.

Finally, the use of  these expressions in the Ricci identity
\begin{eqnarray}
 {x_{\mu ;\nu}}^{ ; \mu}- {(x_\mu^{\ ;\mu})}_{;\nu}&=&  \left(S_{\mu\nu} + \frac{R}{3} \,g_{\mu\nu}\right)x^\mu \,,
 \label{ricci}
\end{eqnarray}
written for the vectors $ s_\mu  $ and $ v_\mu  $, gives us $ \zeta= 0 $  as well as  the relation
\begin{eqnarray}
(\mu^2)^{\,^\prime}\left(1+ \frac{4 \mu^2}{3p}\right)t^\mu R_{;\mu}& = & 0\,.
\label{fineq}
\end{eqnarray}
We recall that $\mu^2 \neq 0 $. Furthermore, using  equations (\ref{al1})  and (\ref{kgfinal1}), it is easy to show that  the expression in the round brackets does not vanish identically. Thus, from equation (\ref{fineq}) (see also (\ref{sR})  and (\ref{vR})) it follows that the scalar curvature $ R $ is constant. This contradicts to our initial assumption that  $ R \neq const $ and completes the proof  of the statement made above.

A similar analysis shows that this statement holds  for  type $ D_s $ spacetimes as well.  However, in both cases an interesting question  is that  {\it what happens if one drops the condition $ h = 0 $}. It turns out that in this case one can still prove the same statement, provided that the spacetime admits a hypersurface orthogonal Killing vector. On the other hand, the general case with $ h \neq 0 $  becomes very  involved and unclear.

We turn now to  the case $ \mu^2 = 0 $, which is equivalent to
\begin{eqnarray}
p &= & m^2 - \frac{R}{12}\,,
\label{mu0}
\end{eqnarray}
and taking this into account in equation (\ref{al1}), we find that
\begin{eqnarray}
\lambda = m^2\,.
\label{lamm}
\end{eqnarray}
Meanwhile, from the field equation (\ref{kgfinal0}) it follows that  $ {D\hskip -.25truecm \slash\,}{^2} S_{\mu\nu} =  0 $.  Using this in equation (\ref{id22}), we see that $ h = 0 $.  With these results, from equation (\ref{id2}) it follows that
\begin{eqnarray}
C_{\mu\nu}C^{\mu\nu} = 0 \,.
\label{cotinv}
\end{eqnarray}
Substituting  (\ref{cotrepre1}) into this equation, we find the relation
\begin{eqnarray}
3 p^2 a^2 + b^2 = 0 \,,
\label{cotinvrel}
\end{eqnarray}
which implies that both  $ a = 0 $ and   $ b = 0 $. Therefore,
\begin{eqnarray}
C_{\mu\nu} = 0 \,,
\label{cotinv}
\end{eqnarray}
as can be seen from equation (\ref{cotrepre1}). (We recall that $ c $  is gauged out as  $ h = 0 $) . Thus, {\it with  $ \mu^2 = 0 $, or equivalently $\lambda =  m^2  $, all type D solutions of NMG must be conformally flat}.

\section{Solutions with Constant Scalar Curvature}

We shall  now  discuss  all type D solutions with constant scalar curvature, focusing first on the case $ \mu^2 \neq 0 $. These are homogeneous anisotropic solutions which can be formally divided into two categories: Solutions of TMG  origin and solutions of non-TMG origin i.e. those which do not have their counterparts in TMG.

We begin by noting that for $ R=const $, the functions $ h $,   $ g $ and $ f $, as  seen from equations (\ref{h}), (\ref{gdef}) and (\ref{fdef}), vanish. Then, equation (\ref{covart}) can be written in the form
\begin{eqnarray}
t_{\mu;\nu} & = & \frac{a}{3} \,\epsilon_{\mu\nu}^{\ \ \ \rho}\,t_\rho - \frac{2b}{3p} \, s_{(\mu}v_{\nu)}\,.
\label{covart3}
\end{eqnarray}
Similarly, equation (\ref{cotrepre1}) reduces to the form
\begin{eqnarray}
C_{\mu\nu} &=& a S_{\mu\nu}+ b \left( s_\mu s_\nu - v_\mu v_\nu \right).
\label{cotrepreh}
\end{eqnarray}
Contracting now equation (\ref{kgfinal0}) with $ t^\mu t^\nu $ and using equations (\ref{doper2}) and (\ref{cot1}), after some algebraic manipulations, we obtain that
\begin{eqnarray}
\epsilon^{\mu\rho\sigma} t_{\mu}\left(2 p \, a \,t_{\rho ; \sigma} + {C_{\rho}}^{\nu} t_{\nu ; \sigma} \right)
&=& 2 p \mu^2\,.
\label{rel1}
\end{eqnarray}
Substitution expressions (\ref{covart3}) and (\ref{cotrepreh}) into this equation yields the relation
\begin{eqnarray}
a^2 +\frac{b^2}{3 p^2}& = & \mu^2\,.
\label{ab1}
\end{eqnarray}
On the other hand, writing down the Ricci identity (\ref{ricci}) for the vector $ t_{\mu} $ and contracting the result with  $ t^{\nu} $, we find the relation
\begin{eqnarray}
a^2 - \frac{b^2}{ p^2}& = & 9 p - \frac{3}{2} \, R\,.
\label{ab2}
\end{eqnarray}
Since  $ p $  is constant, as can be seen from (\ref{al1}), comparing equations  (\ref{ab1}) and (\ref{ab2}) it is easy to see that  $ a $  and $ b $ are constants as well. Furthermore, with equation (\ref{cotrepreh}) it is not difficult to see that for $ b\neq 0 $, we again arrive at the relations given in (\ref{divvso}).

\subsection{Solutions of TMG origin}

When $ b = 0 $,  the Cotton tensor in (\ref{cotrepreh}) becomes as
\begin{eqnarray}
C_{\mu\nu} &=& a S_{\mu\nu}\,,
\label{cotrepre3}
\end{eqnarray}
with
\begin{eqnarray}
a & = & \pm \mu\,.
\label{a}
\end{eqnarray}
Comparing this equation with  those  given in  (\ref{cot1}) and (\ref{tmgdirac}), and taking into account (\ref{tmgkg}) and  (\ref{kgfinal0}), we deduce that all type D  solutions of TMG  can be mapped into NMG. The resulting solutions, as can be seen from equations (\ref{kgfinal1}) and (\ref{ab2}), are characterized  by
\begin{eqnarray}
p & = & \frac{m^2}{10}+\frac{17}{120}\, R\,.
\label{pnmg}
\end{eqnarray}
It is straightforward to show that the use of this value of $ p $  in  (\ref{al1}) and (\ref{kgfinal1}) results in  the  adjusting relations, earlier obtained in  \cite{ah1}, for mapping of  the TMG solutions into NMG. For these solutions equation (\ref{covart3}) takes the form
\begin{eqnarray}
t_{\mu;\nu} & = & \frac{a}{3} \,\epsilon_{\mu\nu}^{\ \ \ \rho}\,t_\rho =
\frac{a}{3}\,\left(v_{\mu} s_{\nu} - s_{\mu} v_{\nu}\right),
\label{killling1}
\end{eqnarray}
where in the last step we have used the relations in (\ref{orient1}). That is,  the vector  $ t_{\mu} $  is the Killing vector (see also \cite{chow1,gm}).

We now want to give an elegant  classification of these solutions by a dimensional reduction on this Killing vector, where the sign of the scalar curvature of the two-dimensional subspace (the factor space)  plays a crucial role. We define the projection operator onto the factor space perpendicular to $ t_{\mu} $. It is given by
\begin{eqnarray}
{h^{\mu}}_\nu & = & {\delta^{\mu}}_\nu + t^{\mu}t_{\nu}\,,
\label{projoper}
\end{eqnarray}
and $ {h^{\mu}}_\nu  t^{\nu}= 0 $.  We also define the derivative operator $ D_{\mu} $ with  respect to the two-dimensional ``spatial"  metric $ h_{\mu\nu} $, and   the associated Riemann  tensor $ {r^{\mu}}_{\nu \lambda \tau} $. We have
\begin{eqnarray}
D_{\nu} V_{\mu} & = & h^{\lambda}_{\;\mu}  h^{\sigma}_{\;\nu} \, V_{\lambda ; \sigma}\,,~~~{r^{\mu}}_{\nu \lambda \tau} V_{\mu}  =  2 D_{[\tau} D_{\lambda]} V_{\nu }\,.
\label{3dopercur}
\end{eqnarray}
With  these definitions, the relation between the spacetime Riemann tensor  and the Riemann tensor of the factor space is derived in a usual way  (see for instance, Ref.\cite{wald}). After straightforward calculations, we have
\begin{eqnarray}
r_{\mu\nu \alpha\beta} & = & h^{\lambda}_{\;\mu}h^{\tau}_{\;\nu} h^{\rho}_{\;\alpha} h^{\sigma}_{\;\beta}\left(
R_{\lambda\tau \rho\sigma}
+ 2\, t_{\tau;\lambda} \,t_{\rho;\sigma} + t_{\rho;\lambda} \,t_{\tau;\sigma} - t_{\rho;\tau} \,t_{\lambda;\sigma}\right).
\label{riemann2}
\end{eqnarray}
Taking the projection of equations (\ref{rimtoric}) and (\ref{killling1}) onto the factor space and using the result in (\ref{riemann2}),  we find that
\begin{eqnarray}
r_{\mu\nu \alpha\beta} & = &  \left( 2p+ \frac{R}{6}- \frac{\mu^2}{3} \right)\left(h_{\mu\alpha} h_{\nu\beta} - h_{\mu\beta} h_{\nu\alpha} \right)
\,.
\label{riemann3}
\end{eqnarray}
From this expression it follows that the scalar curvature of the factor space is given by
\begin{eqnarray}
r &= & \frac{21}{20}\,R-\frac{m^2}{5}\,,
\label{ricciscal2}
\end{eqnarray}
where we have  also used equations (\ref{kgfinal1}) and (\ref{pnmg}).

Next, we need to find the covariant  derivatives of the  basis vectors $ s_{\mu} $ and  $ v_{\mu} $. Using equation (\ref{killling1}), together with the properties of the orthonormal basis in (\ref{ortbasis}), it is not difficult to show that the most  general expressions for these derivatives are given by
\begin{eqnarray}
\label{scovar1}
s_{\mu;\nu} & = &  -\,\frac{a}{3} \; t_{\mu} v_{\nu} + A  v_\mu t_\nu + B v_\mu s_\nu + C v_\mu v_\nu
\,,\\[2mm]
v_{\mu;\nu} & = &   \,\frac{a}{3} \; t_{\mu} s_{\nu} - A  s_\mu t_\nu - B  s_\mu s_\nu - C  s_\mu v_\nu\,,
\label{vcovar1}
\end{eqnarray}
where the functions $ A $,  $ B $  and  $ C $ need to be determined. Projecting these expressions onto the factor space, by means of (\ref{projoper}), and comparing the result, for certainty,  with that obtained  for the metric of  two-dimensional hyperbolic space
\begin{eqnarray}
ds_{2}^2 &= & \frac{1}{k^2}\,\left(d\theta^2 + \sinh^2\theta \,d\phi^2\right),~~~~ k^2= -\, \frac{r}{2}\,,
\label{2dmetric}
\end{eqnarray}
we find that
\begin{eqnarray}
 B &= &  0\,,~~~~ C =   k \coth\theta\,.
\label{bc}
\end{eqnarray}
Substitution of these quantities in equations (\ref{scovar1}) and  (\ref{vcovar1}) yields
\begin{eqnarray}
{s_\mu}^{;\mu} = C \,,~~~~{v_\mu}^{;\mu} = 0 \,.
\label{divvso1}
\end{eqnarray}
On the other hand, using  the Ricci identity (\ref{ricci}) for the vector $ v_{\mu} $, with  equation (\ref{divvso1}) in mind, and  contracting the result with  $ v^{\nu} $, we find that $ A= - a/3 $. Thus, equations (\ref{scovar1}) and (\ref{vcovar1}) take their final forms given by
\begin{eqnarray}
\label{scovar2}
s_{\mu;\nu} & = &  -\,\frac{a}{3}\left( t_{\mu} v_{\nu} +  \, v_\mu t_\nu\right) + C  v_\mu s_\nu
\,,\\[2mm]
v_{\mu;\nu} & = &   \,\frac{a}{3} \left(t_{\mu} s_{\nu} +  \, s_\mu t_\nu \right)- C  s_\mu s_\nu \,.
\label{vcovar2}
\end{eqnarray}
It is  not difficult to show that the associated Lie brackets are given by
\begin{eqnarray}
\left[s, t\right] & = & 0\,,~~~ \left[v, t\right]= 0 \,,~~~\left[v, s\right]= - \frac{2a}{3}\,t + C v\,.
\label{liebr1}
\end{eqnarray}
It is convenient now to choose a coordinate system in which the Killing vector is given as $ t = \partial_\tau $.  Then, the use of  (\ref{2dmetric}) and  (\ref{liebr1})  enables us to specify  the remaining  vectors in the form
\begin{equation}
s  =  k \,\partial_\theta \,,~~ v = \frac{k}{\sinh\theta} \left(\partial_\phi - \frac{2 a }{3k^2} \cosh\theta\, \partial_\tau \right)\,.
\label{specsv}
\end{equation}
Meanwhile,  the corresponding dual one-forms result in the spacetime metric
\begin{eqnarray}
ds^2 &= & -\left(d\tau + \frac{2 a }{3k^2} \cosh\theta \,d\phi \right)^2
+ \frac{1}{k^2}\,\left(d\theta^2 + \sinh^2\theta \,d\phi^2\right).\nonumber\\
\label{3dmetric1}
\end{eqnarray}
We see that this metric has an apparent  singularity at $ k=0 $. However, by redefining the coordinates as
\begin{eqnarray}
\tau \rightarrow \tau - \frac{2 a }{3k^2}\,\phi\,,~~~~\theta \rightarrow k\theta\,,
\label{redefcoor}
\end{eqnarray}
one can put it in the form
\begin{eqnarray}
ds^2 &= & -\left(d\tau + \frac{4 a }{3k^2} \sinh^2\frac{k\theta}{2}\,d\phi \right)^2
\nonumber \\[2mm]  & &
+ \,d\theta^2 + \frac{\sinh^2k \theta}{k^2} \,d\phi^2,
\label{3dmetric2}
\end{eqnarray}
which is suitable for taking the limit $ k\rightarrow 0  $.

It is also worth noting that in terms of  the left invariant one-forms of SU(1,1), parameterized  by  the Euler angles,
\begin{eqnarray}
\sigma_1 &= & - \sin\psi \,d\theta + \cos\psi\, \sinh\theta\, d\phi\,,
\nonumber \\[2mm]
\sigma_2 &= &  \cos\psi \,d\theta + \sin\psi\, \sinh\theta\, d\phi\,,
\nonumber \\[2mm]
\sigma_3 &= & d\psi +  \cosh\theta\, d\phi\,,
\label{3leftforms}
\end{eqnarray}
and with $ \tau = (2 a/3k^2)\, \psi $, the metric in (\ref{3dmetric1}) takes its most simple form given by
\begin{eqnarray}
ds^2 &= & -\left( \frac{2 a }{3k^2}\right)^2 \sigma_3^2
+ \frac{1}{k^2}\,\left(\sigma_1^2 + \sigma_2^2\right).
\label{3dmetric3}
\end{eqnarray}
In this case,  instead of (\ref{liebr1}), we have the following  Lie algebra
\begin{eqnarray}
\left[s, t\right] & = & - \frac{3r}{4a}\, v\,,~~ \left[v,  t\right]= \frac{3r}{4a} \, s \,,~~\left[ v,  s\right]= - \frac{2a}{3}\,t \,.
\label{liebr2}
\end{eqnarray}
We recall that the quantities  $ a $  and $ k^2 $  are as given in (\ref{a}) and (\ref{2dmetric}) and they are determined by equations (\ref{ricciscal2}) and (\ref{pnmg}) together with  (\ref{al1}) and (\ref{kgfinal1}). We see that, in general,  depending on the sign of the scalar curvature $ r $ of the two-dimensional factor space, we have three possible spatial geometries. Namely, a sphere for $ r > 0 $, a hyperboloid for $ r <  0 $ and a flat space for $ r = 0 $. Accordingly,  the associated spacetimes  are  homogeneous anisotropic solutions: (i)  the solution  of Bianchi type $ IX $, or with $ SU(2) $ symmetry, (ii)  the solution  of Bianchi type $ VIII $, or  with $ SU(1,1) $ symmetry, (iii) the solution of Bianchi type $ II $, or with the symmetry of the  Heisenberg group.

We note that the type $ D_s $ counterpart of spacetime (\ref{3dmetric3}) can be obtained  by taking $ t\rightarrow  i s $  and $ s\rightarrow i t $ in the above expressions (see also Eqs.(\ref{riccican}) and (\ref{riccicans})). In this case,  Bianchi type $ II $ remains unchanged, whereas Bianchi types $ VIII $ and $ IX $ goes over into each other.

\subsection{Solutions of non-TMG origin}

We turn now to the case of  $ b\neq 0 $ in equations (\ref{covart3}) and (\ref{cotrepreh}). This results in  solutions which are only inherent in NMG. Clearly, in this case the vector $ t_\mu  $  is no longer the Killing vector, but it is still divergence-free. Furthermore, as we have mentioned above, the vectors $ s_\mu  $ and $ v_\mu  $ are divergence-free as well. With these in mind, using equation (\ref{covart3}) and the properties of the orthonormal basis vectors given in  (\ref{ortbasis}), we find  that
\begin{eqnarray}
\label{scovar3}
s_{\mu;\nu} & = &  -\,\frac{1}{3} \left(a + \frac{b}{p}\right)
\; t_{\mu} v_{\nu} + \psi  v_\mu t_\nu
\,,\\[2mm]
v_{\mu;\nu} & = &  \frac{1}{3} \left(a - \frac{b}{p}\right) t_{\mu} s_{\nu} - \psi s_\mu t_\nu \,,
\label{vcovar3}
\end{eqnarray}
where $ \psi $ is a scalar function. Using these relations in the Ricci identity (\ref{ricci}) for the vectors $ s_\mu  $ and $ v_\mu  $, and taking into account equation (\ref{divvso}), we find that
\begin{eqnarray}
\psi &=& 0\,,~~~~~~ p = - \frac{R}{3}\,.
\label{vsetap}
\end{eqnarray}
Next,  we calculate the action of the operator $ {D\hskip -.25truecm \slash} \,$  on the Cotton tensor in (\ref{cotrepreh}), by  means of equations (\ref{cot1}), (\ref{scovar3}) and  (\ref{vcovar3}). Then, comparing the result with  equation (\ref{kgfinal0}), we find that $ a = 0 $, whereas   $ b $ is given by the same relation as in  (\ref{bvalue}). (See also Eq.(\ref{ab1})).

On the other hand, from equations (\ref{al1}), (\ref{kgfinal1}) and (\ref{ab2}) it follows that
\begin{eqnarray}
p & = & - \frac{4}{15}\, m^2\,,~~~~~~ \lambda  = \frac{m^2}{5}\,.
\label{cosmassrel}
\end{eqnarray}
Finally, using these results in equations (\ref{covart3}), (\ref{scovar3}) and  (\ref{vcovar3}) it is straightforward  to show that  we have the following Lie algebra
\begin{eqnarray}
\left[s, t\right] & = &  \sqrt{2/5}\, m \,v
\,,~~~ \left[v,  t\right]=  \sqrt{2/5}\, m \,s \,,~~~\left[ v,  s\right]=0\,.
\label{liebr3}
\end{eqnarray}
It is convenient to choose a coordinate system in  which the Lie algebra admits the following representation for the basis vectors
\begin{eqnarray}
t&=& \partial_\tau\,,~~~~~ s = \frac{1}{\sqrt{2}}\left(e^{-\sqrt{2/5}\, m \tau} \partial_x + e^{\sqrt{2/5}\, m \tau}  \partial_y\right),
\nonumber \\[3mm]
v &= &\frac{1}{\sqrt{2}}\left(e^{-\sqrt{2/5}\, m \tau} \partial_x - e^{\sqrt{2/5}\, m \tau} \partial_y\right).
\label{spectsv}
\end{eqnarray}
Then, the spacetime metric can easily be written down by using  the associated dual basis. We have
\begin{eqnarray}
ds^2 &=& -d\tau^2 + e^{2 \sqrt{2/5}\,m \tau}\,dx^2 + e^{-2\sqrt{2/5}\,m \tau}\, dy^2\,.
\label{metricb6}
\end{eqnarray}
This solution, as can be seen from the Lie algebra in (\ref{liebr3}), is of a homogeneous anisotropic spacetime of Bianchi type $ VI_{0} $, or with $ E(1,1) $  symmetry.

Again, the corresponding type  $ D_s $  solution can be  obtained by making  the replacement  $ t\rightarrow i s $  and $ s \rightarrow i t $. In this case, we have  the Lie algebra
\begin{eqnarray}
\left[t, s\right] & = &  - \sqrt{2/5}\, m \,v
\,,~~~ \left[v,  s\right]=  \sqrt{2/5}\, m \,t \,,~~~\left[ v,  t\right]=0\,,
\label{liebr4}
\end{eqnarray}
which enables us to specify the basis vectors in the form
\begin{eqnarray}
s &=& \partial_x,,~~~ v= \sin( \sqrt{2/5}\,m x) \,\partial_\tau + \cos( \sqrt{2/5}\,m x)\, \partial_y\,,
\nonumber \\[3mm]
t&= & - \cos( \sqrt{2/5}\,m x)\, \partial_\tau  + \sin( \sqrt{2/5}\,m x)\,\partial_y\,.
\label{spectsv1}
\end{eqnarray}
The associated dual basis results in the solution given by
\begin{eqnarray}
ds^2 &=& \cos (2 \sqrt{2/5}\,m x) \left(-dt^2+ dy^2\right)+ dx^2
+ 2\sin(2 \sqrt{2/5}\,m x)\, dt dy \,.
\label{metricb7}
\end{eqnarray}
This is a homogeneous anisotropic spacetime of Bianchi type $ VII_{0} $, or with $ E(2) $  symmetry.

To complete this subsection, we consider now type D solutions with constant scalar curvature, which correspond  to the case $ \mu^2 =0 $. For these solutions, we have the special relation between the cosmological and mass parameters, as given in (\ref{lamm}), and the Cotton tensor in (\ref{cotrepreh}) vanishes as both $ a $ and $ b $ are zero. Using these facts in equation (\ref{covart3}), we see that
\begin{eqnarray}
t_{\mu;\nu} &= & 0\,.
\label{tcov0}
\end{eqnarray}
It follows that $ t = \partial_\tau $ is a hypersurface orthogonal Killing vector of constant length. With this in mind, we employ the Ricci identity (\ref{ricci}) for the vector $ t_{\mu} $ and obtain that
\begin{eqnarray}
p &= & \frac{R}{6}\,.
\label{pmu0}
\end{eqnarray}
Using this value of $ p $ in equations (\ref{mu0}) and (\ref{riemann3}), we find that
\begin{eqnarray}
R &= & r= 4 m^2\,.
\label{Rr}
\end{eqnarray}
Clearly, type $ D_t $ spacetime metric can now be written as
\begin{eqnarray}
ds^2 &= & -d\tau^2
+ \frac{1}{k^2}\,\left(d\theta^2 + \sinh^2\theta \,d\phi^2\right),
\label{mu0metric1}
\end{eqnarray}
where
\begin{eqnarray}
 k^2= - 2 m^2= -2\lambda \,.
\label{kk}
\end{eqnarray}
Meanwhile, type $ D_s $  solution  is obtained  from this metric by performing   the coordinate changes  $ \tau \rightarrow i \phi $ and  $ \phi \rightarrow  i \tau $. We have
\begin{eqnarray}
ds^2 &= & d\phi^2
+ \frac{1}{k^2}\,\left(d\theta^2 - \sinh^2\theta \,d\tau^2\right).
\label{mu0metric2}
\end{eqnarray}
It is straightforward to show that this metric is locally isometric to the Kaluza-Klein solution  $ AdS_2\times S^1 $  found earlier in \cite{clement} (see also Ref.\cite{berg}). We note that the spacetime  metric (\ref{mu0metric1})  has also  a ``cousin" with   spherical  spatial section (the flat case is trivial). One can show that the type $ D_s $ counterpart of the latter is locally isometric to the   $ dS_2\times S^1 $ solution of \cite{berg}. We recall that similar solutions with a hypersurface orthogonal Killing vector absent in TMG due to the no-go theorem of \cite{an}.

\section{Solutions with Nonconstant Scalar Curvature}

Now we discuss all  type D solutions which possess nonconstant scalar curvature. We recall that these are conformally flat solutions for which $ C_{\mu\nu}=0 $ ($ a=b=c=0 $). Using the traceless Ricci tensor (\ref{riccican})  in the Bianchi identity (\ref{contrbian}) and contracting the result with the vector $ t^\mu $,  we obtain that
\begin{eqnarray}
{t_\mu}^{;\mu} = 0 \,,
\label{divtfin1}
\end{eqnarray}
which  in turn implies  that $ g = 0 $. In obtaining  these results we have  also used equations  (\ref {gdef}) and (\ref{mu0}). With these in mind, from equation (\ref{covart}) we have
\begin{eqnarray}
t_{\mu;\nu} &= &  \frac{v^\rho p_{;\rho}}{p}\,\,v_\mu t_\nu\,.
\label{tcovf1}
\end{eqnarray}
In what follows, we will restrict ourselves to the case when the spacetime under consideration admits at least  one Killing vector $ \xi $. For further convenience, we use the invariance of metric (\ref{metric}) with respect to rotations in the $( s_\mu, v_\mu )$-plane and represent the Killing vector
in the form
\begin{eqnarray}
\xi_{\mu} &= &  \alpha  t_\mu + \beta s_\mu\,,
\label{kilrep1}
\end{eqnarray}
where $\alpha $ and  $ \beta $ are scalar functions to be specified later. Then, due to the  same rotations  we have
\begin{eqnarray}
t_{\mu;\nu} &= &  \frac{v^\rho p_{;\rho}}{p}\,\,v_\mu t_\nu + \frac{s^\rho p_{;\rho}}{p}\,\,s_\mu t_\nu\,,
\label{tcovf2}
\end{eqnarray}
instead of  (\ref{tcovf1}). From the Bianchi identity (\ref{contrbian}) it follows that
\begin{eqnarray}
\xi^\mu {S_{\mu\nu}}^{ ;\nu}= 0\,.
\label{biakil1}
\end{eqnarray}
Substituting equation (\ref{riccican}) in this expression and using  the relation (\ref{divtfin1}),  we find that
\begin{eqnarray}
(\alpha p)_{;\mu} t^\mu & = & 0\,.
\label{alphap1}
\end{eqnarray}
On the other hand, the use of the fact that
\begin{eqnarray}
\pounds_{\xi} S_{\mu \nu} &= &  0\,,
\label{lie1}
\end{eqnarray}
where $ \pounds_{\xi} $ is the Lie derivative along the Killing vector, gives us the relations
\begin{eqnarray}
(\alpha p)_{;\mu} s^\mu & = & 0\,,~~~~(\alpha p)_{;\mu} v^\mu  =  0\,.
\label{alphap2}
\end{eqnarray}
From equations (\ref{alphap1}) and (\ref{alphap2}) it follows that
\begin{eqnarray}
\alpha &= &  a_0 p^{-1}\,,
\label{a0}
\end{eqnarray}
where $ a_0 $ is a constant.

Using the properties of the orthonormal basis in (\ref{ortbasis}) along with  expression (\ref{tcovf2}) and the Killing equation
\begin{equation}
\xi_{(\mu ;\nu)} = 0\,,
\label{killingh}
\end{equation}
we obtain the relations
\begin{eqnarray}
\label{scovarfin}
s_{\mu;\nu} & = & \frac{s^\rho p_{;\rho}}{p}\,t_\mu t_\nu - \frac{v^\rho \beta_{;\rho}}{\beta}\,\,v_\mu s_\nu\,,\\[2mm]
v_{\mu;\nu} & = & \frac{v^\rho p_{;\rho}}{p}\,t_\mu t_\nu +  \frac{v^\rho \beta_{;\rho}}{\beta}\,s_\mu s_\nu \,,
\label{vcovarfin}
\end{eqnarray}
and
\begin{eqnarray}
t^\mu \beta_{;\mu} & = & 0\,,~~~~s^\mu \beta_{;\mu}  =  0\,.
\label{tsbeta}
\end{eqnarray}

Next, we need to obtain the determining equations for unknown  functions $ \beta $ and $ p $. They are obtained by employing the Ricci identity (\ref{ricci}) for the basis vectors $ t^\mu $,\, $ s^\mu $  and $ v^\mu $, respectively, as well as  using equations (\ref{tcovf2}), (\ref{scovarfin}) and (\ref{vcovarfin}). After some manipulations, we arrive at the ``oscillator" type equation for $ \beta $,
\begin{eqnarray}
\label{beteq}
v^\mu v^\nu \beta_{;\mu;\nu}- k^2 \beta &=& 0\,,
\end{eqnarray}
as well as  at the following set of equations for $ p\, $,
\begin{eqnarray}
\label{peq1}
s^\mu s^\nu (p^{-1})_{;\mu;\nu} &=&  3 + k^2 p^{-1}-  \frac{v^\rho \beta_{;\rho}}{\beta}\,v^\mu  (p^{-1})_{;\mu}\,, \\[2mm]
\label{peq2}
v^\mu v^\nu (p^{-1})_{;\mu;\nu} &=&  3 + k^2 p^{-1}\,,\\[2mm]
v^\mu s^\nu (p^{-1})_{;\mu;\nu} &=& 0\,.
\label{peq3}
\end{eqnarray}
We note that  $ k^2 $ is the same as that given in (\ref{kk}). It is straightforward to show that the Killing vector in (\ref{kilrep1}) commutes with the basis vector $ v $ in the sense of their Lie bracket. That is,  we have
\begin{eqnarray}
\left[\xi, v\right] & = &  0\,,
\label{liebrxit}
\end{eqnarray}
which can easily be verified by means of equations (\ref{tcovf2}), (\ref{scovarfin}) and (\ref{vcovarfin}). This fact allows us to choose the vectors $ \xi $  and $ v $  as
\begin{eqnarray}
\xi & = &  \partial_z \,~~~~ v = \partial_ x \,
\label{vxi}
\end{eqnarray}
in a coordinate system $( x\,, y\,,z) $. With this in mind, we turn to equation (\ref{beteq}) whose general solution is given by
\begin{eqnarray}
\beta & = & f_1(y)\sinh(k x) + f_2(y)\cosh(k x)\,,
\label{betasol}
\end{eqnarray}
where $ f_1(y) $ and $ f_2(y) $ are arbitrary functions. Alternatively, the use of the reparametrization invariance  of the vector fields  $ \xi  $ and $ v $  with respect to the coordinate freedom $ x \rightarrow  x + g_1(y) $ enables us to reduce  this solution into three different  forms given by
\begin{eqnarray}
\label{betasol1}
\beta & = & \omega (y) \sinh(k x)\,,\\[2mm]
\label{betasol2}
\beta & = & \omega (y) \cosh(k x)\,,\\[2mm]
\beta & = & \omega (y) \,e^{k x}\,,
\label{betasol3}
\end{eqnarray}
where the function $ \omega (y)$ will be specified below.

We turn now to  equations (\ref{kilrep1}) and (\ref{a0}) and consider the  cases $ a_0=0 $ and $ a_0 \neq 0 $ separately.

\subsection{The case $ a_0= 0 $}

In this case, the Killing vector, as seen from equation (\ref{kilrep1}), becomes proportional to the spacelike basis vector $ s $. That is, we have
\begin{eqnarray}
s & = & \frac{1}{\beta} \,\partial_z\,.
\label{srep2}
\end{eqnarray}
Next, using  the relations
\begin{eqnarray}
\left[\xi, t\right] & = &  0\,,~~~~\left[v, t\right]  =   \frac{v^\mu p_{;\mu}}{p}\,t\,,
\label{liebrxitvt}
\end{eqnarray}
where the first relation  follows from equation (\ref{lie1}) and  the second one is obtained by means of equations (\ref{tcovf2}) and  (\ref{vcovarfin}), we fix the timelike basis vector as
\begin{eqnarray}
t & = & p \,\partial_y\,.
\label{trep2}
\end{eqnarray}
With this in mind, from the first relation of (\ref{tsbeta}), we find that  $\omega= const $ in equations (\ref{betasol1})-(\ref{betasol3}). Using now the dual one-forms to the basis vectors in  equations (\ref{vxi}),  (\ref{srep2}) and (\ref{trep2}) we arrive at   the spacetime metric
\begin{eqnarray}
ds^2 & =&  - p^{-2} d\tau^2 + dx^2 + \beta^2 dy^2\,,
\label{metricxi1}
\end{eqnarray}
where we have relabeled the coordinates as  $y=\tau $ and  $ z=y $. We see that this metric in general is characterized by two functions $ \beta = \beta(x) $ and $ p = p(\tau, x ) $. The first function is given by  solutions (\ref{betasol1})-(\ref{betasol3}), whereas the second function  is determined by  the system of differential equations (\ref{peq1})-(\ref{peq3}), which admits the general solution
\begin{eqnarray}
p^{-1} & =& -\frac{3}{k^2} + f(\tau) \partial_x \beta\,,
\label{solk0}
\end{eqnarray}
where $ f(\tau) $ is a smooth function. Performing an analytical continuation of solution (\ref{metricxi1}) by $ \tau \rightarrow i y $ and   $ y \rightarrow i \tau$,  we arrive at its type $ D_s $ counterpart. It is straightforward to show that the resulting spacetime metrics for the explicit forms of $ \beta $ given in (\ref{betasol1})-\ref{betasol3}),  after appropriate coordinate changes, represent the  black hole, gravitational soliton and kink type solutions with one Killing vector, respectively. Meanwhile, for $ f= const $ the second Killing vector appears as well, and these solutions reduce to those with two commuting Killing vectors, found earlier in \cite{ troncoso}. Below, we shall focus only on the black hole type solution with one Killing vector.

Let us take  $ \beta= \sinh (k x) $ in metric (\ref{metricxi1}). Then, passing to the  coordinates  $  3/k^2- r  = M \cosh (\nu x)$, where $ r $ is a radial coordinate and $ M $ is a constant, $\tau \rightarrow i \phi \, $  and  $ y \rightarrow i k M \tau $, we arrive at the  metric
\begin{eqnarray}
ds^2 & =&  - k^2 \left(r- r_{+}\right)\left(r- r_{-}\right)d\tau^2 + \frac{dr^2}{k^2 \left(r- r_{+}\right)\left(r- r_{-}\right)} \nonumber \\[2mm]
&& + \left[r-(r_{+} + r_{-})/2 + F(\phi)\right]^2 d\phi^2\,,
\label{gnbh1}
\end{eqnarray}
where the radii of outer $ (r_{+})$ and inner $ (r_{-})$  horizons are given by \begin{eqnarray}
r_{\pm} & = & \frac{3}{k^2}  \pm M\,,
\label{outinhor}
\end{eqnarray}
and  $ F(\phi)$ is an arbitrary function. This metric can be interpreted as describing  a ``generalized"   black hole type solution, which is asymptotically AdS spacetime with $\Lambda= 2m^2 $ $ (m^2 <0) $. For  $ r_{+}= r_{-} =r_0 $,  it corresponds  to an extremal black hole with one Killing vector, which was earlier described in \cite{ah1}. It is also easy to see that for $ F(\phi)= const $, this  solution  goes over into a standard new black hole metric (with two commuting Killing vectors) of \cite{berg, troncoso}.

In light of this, it is  also interesting to ask  whether  the solution in (\ref{metricxi1}) admits another limit with two  Killing vectors. It turns out that the answer is affirmative, but the Killing vectors are no longer commuting. In fact, solving the equation  $ \eta_{(\mu ;\nu)} = 0 $ for the putative Killing vector $ \eta $, we find that
\begin{eqnarray}
\eta  & = & \partial_\tau - \frac{b_0}{k} \, \partial_x + b_0 y \partial_y \,,
\label{noncomkil1}
\end{eqnarray}
provided that the metric functions are given by
\begin{eqnarray}
f & = & e^{b_0 \tau}\,,~~~~ \beta   =  e^{k x}\,.
\label{metfuncs}
\end{eqnarray}
Here $ b_0 $ is a constant of integration. We see that  the  Killing vectors $ \eta $ and $ \xi $ do not commute for $ b_0 \neq 0 $. Clearly, similar analysis  is also true for the type $ D_s $ counterpart of (\ref{metricxi1}).

\subsection{The case $ a_0 \neq 0 $}

We begin with calculating, in addition to (\ref{liebrxit}), the other Lie brackets of the vectors $ \xi $,  $ s $  and $ v $. Using  equations (\ref{kilrep1}) and  (\ref{scovarfin}) and taking into account equation (\ref{tsbeta}) it is easy to show that
\begin{eqnarray}
\left[\xi, s\right] & = &  0\,,
\label{liebrxis}
\end{eqnarray}
whereas, the use of equations (\ref{scovarfin}) and (\ref{vcovarfin}) yields
\begin{eqnarray}
\left[s, v\right] & = &  \frac{v^\mu \beta_{;\mu}}{\beta}\,s\,.
\label{liebrsv}
\end{eqnarray}
With these Lie brackets  and  with equation (\ref{vxi}), one can show that the vector $ s $ admits the simple representation
\begin{eqnarray}
s & = & \frac{1}{\beta} \,\partial_y\,,
\label{srep1}
\end{eqnarray}
and $\omega= const $ in equations (\ref{betasol1})-(\ref{betasol3}). In obtaining these results we have also used the first equation in (\ref{tsbeta}) and the coordinate freedom $ z \rightarrow  z + g_2(y)\, $.

Next, using the above representation of the vectors $ \xi $ and   $ s $ in equation (\ref{kilrep1}), we see that the timelike basis vector $ t $ is given by
\begin{eqnarray}
t & = & p \left(\partial_z - \partial_y\right),
\label{trep1}
\end{eqnarray}
where we have set $ a_0=1 $, for certainty. It is easy to see that  the associated dual one-forms  enable us to write down the spacetime metric in the form
\begin{eqnarray}
ds^2 & =&  - p^{-2} d\tau^2 + dx^2 +\rho^2 \left(dy+ \omega d\tau\right)^2,
\label{metricxi2}
\end{eqnarray}
where  we have passed to the new coordinates  $ \rho \rightarrow  \beta/\omega  $\,, $ \tau \rightarrow z $ and $ y \rightarrow  y \omega  $. We recall that $ \rho = \rho(x) $ is determined by the solutions in (\ref{betasol1}) - \ref{betasol3}), and $ p = p(x, y)$ is given by the solutions of equations (\ref{peq1})-(\ref{peq3}). Thus, we have the following three pairs of solutions
\begin{eqnarray}
p^{-1}& =& -\frac{3}{k^2} + c_0 \cosh(k x) + \left(c_1  e^{i k y}+  c_2 e^{- i k y}\right)\sinh(k x),\nonumber\\[2mm]
\rho & =& \sinh(k x),
\label{pair1}
\end{eqnarray}
\begin{eqnarray}
p^{-1}& =& -\frac{3}{k^2} + c_0 \sinh(k x) + \left(c_1  e^{k y}+  c_2 e^{-k y}\right)\cosh(k x),\nonumber\\[2mm]
\rho & =& \cosh(k x),
\label{pair2}
\end{eqnarray}
\begin{eqnarray}
p^{-1}& = & -\frac{3}{k^2} + c_0 \,e^{k x} + c_1 \left(k^2 y^2  e^{k x}+  e^{-k x}\right) +c_2\, y  e^{k x}
,\nonumber\\[2mm]
\rho & =& e^{k x},
\label{pair3}
\end{eqnarray}
where $ c_0 \,$, $ c_1 $ and $ c_2 $  are constants of integration. The type  $ D_s $ counterpart of solution (\ref{metricxi2}) is obtained by making the coordinate changes $ \tau \rightarrow i y $ and   $ y \rightarrow i \tau$.

It is important to note that metric (\ref{metricxi2}) with one Killing vector is intrinsically different from that in (\ref{metricxi1}) even for $ \omega= 0 $. In particular, this can be  seen from the fact that for  metric  (\ref{metricxi1}), for which  $ a_0=0 $, the Killing vector in (\ref{kilrep1}) is an eigenvector of (\ref{riccican}), i.e. $ {S_{\mu}}^\nu \xi_\nu= p \,\xi_\nu  $. However, this is not the case for metric (\ref{metricxi2}), where $ a_0\neq 0 $ and we have  $ {S_{\mu}}^\nu \xi_\nu= p \,\xi_\nu - 3 a_0 t_\nu $. On the other hand, for the vanishing constants $ c_1 $ and $ c_2 $  an additional Killing isometry appears in (\ref{metricxi2}), and $ \partial_y $ becomes the second Killing vector, commuting with the first one, $ \xi=\partial_\tau $. With these two commuting Killing vectors, solutions (\ref{metricxi1}) and (\ref{metricxi2}) become locally isometric  to each other.

In analogy with the case of (\ref{metricxi1}), it is not difficult to show that  for the special forms of the metric functions, solution (\ref{metricxi2})  still admits a second Killing vector. This vector does not commute with  the Killing vector $ \xi $. Again, solving the associated Killing equation, we find the explicit form of the Killing vector. It is given by
\begin{eqnarray}
\eta  & = & e^{k(\omega \tau+y)}\left[\tanh(k x) \,\partial_y - \partial_x\right]\,,
\label{noncomkil2}
\end{eqnarray}
provided that  the metric functions are as in equation (\ref{pair2}) with $ c_0=0 $ and $ c_2=0 $. We note that for $ \omega =0 $, this vector commutes with the Killing vector $ \xi $.  Then, it is not difficult to show that the resulting metric with these commuting Killing vectors is locally isometric to that in (\ref{metricxi1}) with $ f= const $ and with $ \beta= e^{k x} $. It is also worth to note that the Killing vector, similar to that in (\ref{noncomkil2}), does also exist for the solutions in (\ref{pair1}), again  with the vanishing $ c_0$ and $ c_2 $. On the other hand, there is no such a Killing vector for the solutions in (\ref{pair3}). Finally, it is certainly interesting to know what is the relation between metrics (\ref{metricxi1}) and (\ref{metricxi2}) in their special limits with the Killing vectors in (\ref{noncomkil1}) and  (\ref{noncomkil2}), respectively. This issue  requires further investigation.

In summary, we note that the spacetime metrics in (\ref{metricxi1}) and (\ref{metricxi2}), both admitting one Killing vector, represent all type D solutions of NMG with nonconstant scalar curvature. We recall once again that all these solutions are conformally flat and they do exist provided that the special relation $\lambda=m^2 $  holds.

\section{Conclusion}

This paper completes our programme of the exhaustive investigation of types  D and N exact solutions of NMG,  which was first begun in \cite{ah1} and then continued in \cite{ah2}. Here we have presented an exhaustive set of  solutions, which includes all type D solutions with both constant and nonconstant curvatures. As in the previous cases, this was achieved in the framework of a novel proposal that amounts to reformulation of NMG in such a way that the field equations underlying the theory acquire a remarkably simple  form. Namely, they reduce to a massive (tensorial) Klein-Gordon type equation which is accompanied by a constraint equation as well. The ``mass term" in this equation plays a distinguished  role in many aspects of the delineation of exact solutions. In particular, for its constant value being achieved only for type N and type D (with constant scalar curvature) spacetimes, there exists an intimate relation between the theories of NMG and TMG; the latter can be understood as the square root of the former, thereby opening up the way for mapping all the associated solutions of TMG into NMG.

We have proved that for the nonzero value of the mass term, all type D solutions of  NMG must have constant scalar curvature, whereas for the vanishing mass term all the solutions turn out to be conformally flat, possessing both constant and nonconstant  scalar curvatures. Introducing an orthonormal  basis  of three real vectors and using the gauge freedoms provided by the associated Lorentz symmetries, we have given a detailed description of type D spacetimes  in NMG, in terms of these basis vectors and their covariant derivatives.

With type D solutions of constant scalar curvature, we have shown that all the solutions fall into two classes: (i) homogeneous anisotropic solutions which consist of those having their counterparts in TMG and those of being only inherent in NMG, (ii) solutions with vanishing Cotton tensor, i.e. conformally flat solutions. Using the dimensional reduction procedure on a Killing vector, we have given an elegant classification of  the homogeneous anisotropic solutions of TMG origin in terms of the scalar curvature of two-dimensional subspace. We have obtained the most compact expression for the spacetime metric and established a {\it universal} Lie algebra for the associated basis vectors. Depending on the value (positive, negative or zero) of the two-dimensional scalar curvature, appearing in the Lie  algebra, there exist three possible spatial geometries and the associated spacetimes  are of Bianchi types $ IX $,\, $ VIII $ and $ II $, respectively. We have also obtained all homogeneous anisotropic solutions of non-TMG origin, which are of Bianchi types $ VI_{0} $ and $ VII_{0} $ spacetimes.  It is important to note that these solutions  require a special relation between the  cosmological and mass parameters $(\lambda=m^2/5 ) $, unlike the  Bianchi types  of TMG origin. As for conformally flat solutions, they exist only for $\lambda=m^2 $ and possess  a hypersurface orthogonal Killing vector of constant length. We have discussed these solutions as well, emphasizing that some of them  are locally isometric to the previously-known  Kaluza-Klein  type $ AdS_2\times S^1 $ or $ dS_2\times S^1 $ solutions.

With type D solutions of nonconstant scalar curvature, we have found two general metrics which admit at least one Killing  vector and comprise the entire set of such solutions. All these solutions are conformally flat and require $\lambda=m^2 $ by their very existence.  We have discussed the special limits of these metrics with two commuting Killing vectors. In the latter case, the  resulting  metrics become locally isometric to each other, recovering all the previously-known solutions in the literature. We have shown that there also exists a special limit of both general metrics when two noncommuting Killing vectors appear. The question of whether in this case the resulting  metrics are locally isometric or not is  unclear and it requires further investigation. As an illustrative example, we have briefly discussed global properties of the solutions with  at least one hypersurface orthogonal Killing vector, focusing on black hole type solutions.

In this paper, we have given all the solutions in their local form. It is certainly of great interest to perform the complete global analysis  of all type D solutions presented here, especially in the case with nonconstant scalar curvature. It is also of great interest to extend the results of this paper to other theories of 3D massive gravity. Using our approach, this can be done for ``general massive gravity", for an extension of NMG by  adding a parity violating Chern-Simon term, as well as for the theory of NMG with higher-order curvature invariants. We hope to return to these issues in our future works.

\end{document}